\documentclass{aa}  
\usepackage{graphicx}
\usepackage{epsfig}
\usepackage{textcomp}

\usepackage{gensymb}
\usepackage{txfonts}
\usepackage{subfigure}
\usepackage{amsmath}
\usepackage{gensymb}

\usepackage{xcolor}
\usepackage{listings}

\begin{document} 
   \title{Identification of activity peaks in time-tagged data with a scan-statistics
     driven clustering method and its application to gamma-ray data
     samples$^1$}

   \titlerunning{A scan-statistics driven clustering method for gamma-ray flare studies}

   \author{L. Pacciani \inst{1} }

   \institute{Istituto di Astrofisica e Planetologia Spaziali - Instituto Nazionale di Astrofisica (IAPS-INAF),
              Via Fosso del Cavaliere, 100 - I-00133 Rome (Italy)\\
              \email{luigi.pacciani@iaps.inaf.it} }

\authorrunning{L. Pacciani \inst{1}}
  \abstract
      {The investigation of activity periods  in time-tagged data samples is a
        topic of large interest. Among Astrophysical samples, gamma-ray
        sources are widely studied,
        due to the huge quasi-continuum data set available
        today from the FERMI-LAT and AGILE-GRID gamma-ray telescopes.}        
   {To reveal flaring episodes of a given gamma-ray source, researchers make use of
     binned light-curves. This method suffers several drawbacks: the results depends on
     time-binning, the identification of activity periods is difficult
     for bins with low signal to noise ratio.
     A different approach is investigated in this paper.}
   { I developed a  general temporal-unbinned method to identify flaring
     periods in  time-tagged data
     and discriminate statistically-significant
    flares: I propose an  event clustering method in one-dimension to identify flaring episodes,
     and Scan-statistics to evaluate the flare significance within the
     whole data sample.
     This is a photometric algorithm. The comparison of the
     photometric results (e.g., photometric flux, gamma-ray spatial distribution) for the
     identified peaks with the standard likelihood analysis for the same
     period is mandatory to establish if source-confusion is spoiling results.
   }
   {
    The procedure can be applied to reveal flares in any time-tagged data sample.
    The study of the gamma ray activity of 3C 454.3 and of the fast variability of the Crab Nebula are shown as examples.
     The result of the proposed method is similar to a photometric
     light curve, but peaks are resolved,  they are statistically significant
     within the whole period of investigation, and peak detection capability does not suffer time-binning
     related issues.\\
     The method can be applied
     for gamma-ray sources of known celestial
     position, for example, sources taken from a catalogue.
     Furthermore the method can be used when it is
     necessary to assess the statistical significance within the whole
     period of investigation of a flare from an unknown gamma-ray source.
}
   {}

 
   \maketitle

\addtocounter{footnote}{1}   
   \footnotetext{C code implementing the whole procedure and supplementary
     material is available in electronic form
     at the CDS via anonymous ftp to \url{cdsarc.u-strasbg.fr} (130.79.128.5)
     or via \url{http://cdsweb.u-strasbg.fr/cgi-bin/qcat?J/A+A/}.}

%

\section{Introduction}

  The study of  source variability is a fundamental topic
  in Astrophysics. I treat here the problem of peak activity identification in time-tagged data samples.
  This study is primary motivated by the huge Fermi-LAT and AGILE archival
  data sample.\\
  The Fermi Satellite operations allow for each position of the sky to be
  observed every orbit or two, and in some case every 4 orbits of the
  satellite around the Earth.
  This strategy is operating for the entire observation period (9 years, up
  to now) of the FERMI-LAT pair production gamma-ray telescope \citep{atwood2009}.\\
  AGILE \citep{tavani2009} also operates in spinning mode and scan the whole sky every $\sim$7 minutes.\\  
  These scanning-sky strategies allow researcher to study gamma-ray variability of Celestial Sources
  for the whole Missions lifetime, without gaps.\\
  The two preferred methods of investigation are likelihood analysis
  \citep{mattox1996} and binned light-curves.\\
  The Likelihood analysis,  can be applied for a fixed integration period,
  such as for the preparation of a catalog, or for the observation of a field within a predefined period.\\
  The binned light-curve analysis is the preferred method to investigate
  source variability. It introduces an obvious timescale (the
  time-bin). Several light curves (varying  time-bin and bin positions) are usually
  applied to data to recognize flares, their duration, and peak activity.
  The statistical significance for every bin of the light curves
  is evaluated. It refers to the significance of source signal against background
  within the chosen time-bin.
  False flare detection post-trial probability ($P_{false}$) can be
  evaluated \citep[see, e.g.,][]{bulgarelli2012} in order to study
  significance of the flaring activity period against background statistical
  fluctuations during the whole period of investigation.
  The evaluated $P_{false}$ can be used when mean source signal is negligible with respect to background
  within the whole scrutinized period.\\
  \cite{lott2012} proposes an adaptive light curve strategy to overcome the
  drawbacks arising from a fixed time-bin.\\
  An other method to study gamma-ray source variability from the FERMI-LAT data sample
  is investigated in \cite{ackermann2013}.  \\
  
  Scan-statistics  \citep{nauss1965}  evaluates statistical significance
  of the maximum number of events found within a moving window of fixed  length
  \citep[see][for a review]{wallenstein2009}. 
  It is the natural approach to study the statistical
  significance of detection of non-random clusters against the hypothesis of an uniform distribution on an
  interval.
  \cite{nauss1966} obtained that the power of scan-test is larger
  with respect to disjoint-test (e.,g., with respect to binned light curve
  investigation) assuming to know the cluster size.
  \cite{nagarwalla1996,glaz2006} investigated the case of an
  unknown cluster size, but the results of their methods depend on several hypothesis.\\
  \cite{cucala2008} studied an hypothesis-free method based on the distance
  $D_{i,j}\ = \ X_{(j)}\ -\ X_{(i)}$ among ordered events $i,j$ where $X_{(i)}$ is the
  position of the event of index $i$ in the extraction interval.\\
  I propose here a similar clustering method to identify activity periods above
  an assumed uniformly distributed data sample.
  It allows to obtain unbinned light curves of astrophysical sources
  for time-tagged data samples, and  it overcomes time-bin related drawbacks.\\ 
  It is a photometric method (e.g., the flux is evaluated within an extraction radius)
  and, contrary to likelihood based analysis, it lacks the simultaneous study of nearby sources.\\
  Its worth mentioning an other algorithm  developed to
  produce unbinned light curves and show structures within the flare profiles, the Bayesian Block
  \citep[see][and reference therein]{scargle1998,scargle2013}.\\  
  The procedure proposed in this paper is able to resolve candidate flares with
  a flux (including background) larger than the mean flux and background within
  the examined period.\\
  In this paper I explain the method following the study case of Fermi-LAT gamma-ray data sample.
  The generality of the method shows up anyway.\\   
  In the following sections, I will present a concise description of the method (section 2),
  the details of clustering method
  (section 3 and 4),
  Scan-Statistics applied to the problem (section 5 and 6),
  the construction of {\em unbinned light curves} to identify activity peaks
  (section 7). All the sections listed above explain the method in the
  general case of an ordered set of events.\\
  The Monte Carlo study of the
  procedure and the performance for the Fermi-LAT Telescope will be shown in section 8.\\
  In section 9 I will show examples of application to gamma-ray
  sources, and I also discuss drawbacks arising with the method.\\
  The proposed method is photometric. The obtained candidate flares
  could suffer source spatial confusion. The comparison with the full likelihood
  analysis is mandatory to assess the level of source-confusion, and
  eventually reject questionable flares.\\
  Section 10 summarizes, performance and weakness of the method  for the
  generic case of ordered data samples.\\

\section{Step-by-step definition of the algorithm}
\label{section:coincise_step_by_step}
The proposed algorithm is applied to an ordered sample of events of size $N$.\\
It consists of: a method of event clustering iterated to obtain all the
conceivable clusters from the original sample; an ordering procedure among clusters; and a
Statistical evaluation of non-random clusters. 
The entire method depends on  2 parameters:
the statistical Confidence Level (c.l.) and the $N_{tol}$ parameter (its meaning will
be explained in section 3, and in appendix \ref{section:fragmentation_and_ntol}
it is explained how to correctly choose it).\\

Clustering method: The clustering method has two parameters ($N_{tol}$ and
$\Delta_{thr}$). The $N_{tol}$ parameter is kept constant. $\Delta_{thr}$ naively corresponds to the maximum allowed
distance among the elements belonging to a certain cluster.\\ 
The clustering procedure is iterated scanning on the $\Delta_{thr}$ parameter.
It is finely decreased starting from the largest spacing among contiguous
 events of the data sample under investigation.
The $\Delta_{thr}$ decreasing procedure stops when only clusters of size 2 (of two events) remain. At the
end of the scanning, the $\Delta_{thr}$ space is fully explored.\\
As far as we can obtain the same cluster from a sub-set of $\Delta_{thr}$,
cluster duplicates are removed. No more than $N$ clusters are identified.\\

For a generic cluster $C_i$, there are several useful quantities:
Cluster size ($N_i$) is defined as the number of events contained in the
  cluster. Cluster length ($l_i$) is the distance from the first to the last
  event of the cluster. It's also useful to define the effective cluster
  length  ${\tilde l}_i\ :=\ l_i\cdot(1+\tfrac{1}{N_C})$.
  The event density ($\rho_i$) is defined as
  $\rho_i\ =\ \frac{N_i-1}{l_i}$. The cluster boundaries are the position of
  the first and of the last element of the cluster within the extraction
  interval. \\

Ordering: I will show that due to cluster definition, the entire set of
clusters ($\{C_i\}$) is a single-root tree (as defined in set theory).
A set is a single-root tree if an ordering law  exists, such that
for each element A (except for the root element) of the set, there exists a well-ordered sub-set of elements
(the concept of well-ordered set of events is defined in set theory).
For the built set of clusters, $\Delta_{thr}$ is the order parameter, and the
ordering law is the comparison of the order parameter among the clusters.\\
Within a single-root tree, ancestors/descendant and parent/sons relations
are defined. Branches, chains, leaves are defined as well. I will show
  that the boundaries of a cluster are within the boundaries of other clusters
  (ancestor/descendant relationship), or the clusters are disjoint.
The largest cluster of the set contains all the events. I will denote it
{\em  ground cluster}.\\

In this paper clusters are denoted with $C_i$, where $i$ is the positional
index within the ordered set of clusters. Sometimes the notation
$C_i^l$ is used for the same cluster. In this case, the upperscript is the positional index of the
parent of $C_i$.\\

Removal of random clusters:\\
It is based on the evaluation of Statistical significance of a son cluster, starting from the
null hypothesis that the events of its parent are uniformly distributed.\\
A multiple window Scan-Statistics based method is used to assess the probability to obtain a
cluster by chance starting from the parent cluster.\\
Once the ordered set $\{C_i\}$ has been prepared, the removal of random clusters is performed starting from ground cluster, and
ascending the tree $\{C_i\}$ (e.g., moving in the direction of lowering $\Delta_{thr}$). At  first the statistical significance of the
ground cluster is evaluated, with the null hypothesis that the events of the whole observing period are
uniformly distributed. If the null hypothesis is accepted (according to the chosen
confidence level), the ground cluster is
removed, otherwise it is maintained within the set of clusters.\\
For each cluster $C_*$ that has to be evaluated, its parent is identified within the ordered set. The
null hypothesis is that the events within the parent cluster are uniformly distributed.
If no parent exists within the set of clusters, the null hypothesis is that the events within the whole
observing period are uniformly distributed. If the null hypothesis is
accepted, then $C_*$ is removed; otherwise it is kept.\\

It is a convenient choice to add a special object as first element of
$\{C_i\}$. Its boundaries are the start and stop of the whole observing
  period; its size is the size N of the whole sample; its event density is 
$\rho_{whole}=\tfrac{N}{L}$, where $L$ is the duration of the whole observing
  period.
This special object is the {\em root} of the tree, and it does not obey to the
chosen cluster condition.\\
After removal of random clusters, Two scenarios occur:
no clusters  remain apart the root
(steady source activity), and  the root describes the steady case. The
  opposite case is that some clusters survive the removal procedure. 
The remaining clusters still form a tree. The leaves of the tree are the
activity peaks, and chains connecting the {\em root} to the leaves describe flaring periods.\\  
The temporal diagram describing the full tree of survived clusters is called {\em unbinned light-curve}.

\section{Clustering Method} \label{section:clustering_method}
  A clustering method is applied to an ordered sample of events.
  The case of Fermi-LAT data is useful to understand, and to apply the procedure:\\  
  Suppose We want to study the variability of a gamma-ray source with known coordinates. In this
  case, the ordered sample is identified with gamma-ray events recorded within
  a chosen extraction region centered on the source coordinates.
 The extraction region is chosen according to the instrumental {\em point spread function} (PSF).
 As far as the PSF for the event $i$ depends on the reconstructed energy
 ($E_i$) and on the morphology of the reconstructed $e^-e^+$ tracks ($type_i$), 
 I choose the radius of the extraction region coincident with the 68\% containment radius at the given
 energy and the given event type. I denote the extraction region of the event
 $i$ with $R_{68}(type_{i},E_{i})$.\\  

  The Fermi-LAT exposure to each position of the sky rapidly changes with
  time, as the satellite scan the whole sky within a few orbit. The
  cumulative exposure ($\xi_i$, defined as the exposure from the start of the FERMI-LAT
  operation to the time of the i$^{th}$ gamma-ray to be scrutinized) 
  is a convenient domain  to build the clustering. In facts, for steady
  sources, the expected number of collected gamma-rays in a time interval is proportional to
  the exposure of the interval. The time-domain, instead lacks that property.\\
  
  For each source, The data set is the cumulative exposure $\xi_i$
  of the gamma-ray events collected within $R_{68}(type_{i},E_{i})$ within the chosen
  observing period.

  The data set of the cumulative exposure is denoted with  $\{\xi_i\}$, and
  the ordered set is denoted with  $\{\xi_{(i)}\}$.\\
    Here, and after in the paper,
    the generalization of the problem is easily performed, considering the
    generic ordered set of observables $\{x_{(i)}\}$ (which  are supposed to
    be uniformly distributed) instead of  $\{\xi_{(i)}\}$.\\

 \cite{cucala2008} proposed to define clusters starting from the {\em distance}
  $D_{i,j}\ = \ x_{(j)}\ -\ x_{(i)}$ for all the ordered events $i,j$.\\
  A different definition of a cluster is the following:
  Events form a cluster if the relative distance (in the exposure-domain) from
  each event to the previous one is less than a specified threshold ($\Delta_{thr}$):\\
  \begin{equation}
    \xi_{(i)} -\xi_{(i-1)} \ < \ \Delta_{thr} \ . \label{eqn:cluster1}
  \end{equation}
  with this definition, a period of steady flux \citep[e.g., for a steady source, or during an activity period
  with a plateau, such as was reported for  3C 454.3 on the first half of
  November 2010, see ][]{abdo2011b} can be fragmented in two or more clusters, due to a
  peculiar spacing of events:
  In such a case, when a $\Delta^*_{thr}$ is chosen, such that $1/\Delta^*_{thr}$ corresponds to the mean source
  flux within the period (F$_{flat}$), the probability for each photon to stay at a
  distance $>\Delta^*_{thr}$ from the previous one is $1/2$. Hence, the flat flux period
  corresponds to a large number of clusters (fragmentation,
  see appendix \ref{section:fragmentation_and_ntol} for a definition of fragmentation).\\

  To avoid fragments, the definition of a cluster must be changed.\\
Clusters are defined as the largest uninterrupted sequences of contiguous events  of the ordered set $\{\xi_{(i)}\}$, such that
for each event $l$ of each sequence, there exist the elements  $i$ and $i+k$ of $\{\xi_{(i)}\}$
for which the following conditions are satisfied:
\begin{equation}
  \label{eqn:cluster2} 
\left\{
\begin{aligned}
  l \ \in \ [i,i+k] \\
  \xi_{(i+k)}\ -\ \xi_{(i)} & < &  k\cdot \Delta_{thr}   \ \  (\ k\ \leq \ N_{tol} \ ) \\
\end{aligned}
\right.  
\end{equation}
where N$_{tol}$ is a tolerance parameter for the cluster definition. 
The clustering scheme reported in   eq. \ref{eqn:cluster2} is called {\em
  short range search} (SRS) clustering scheme 
\footnote{The C language code to identify clusters according to eq.
  \ref{eqn:cluster2} is available in electronic form
  at the CDS via anonymous ftp to \url{cdsarc.u-strasbg.fr} (130.79.128.5)
  or via \url{http://cdsweb.u-strasbg.fr/cgi-bin/qcat?J/A+A/}}.
  Eq. \ref{eqn:cluster2} is forward-backward symmetric: substituting $i$ with $i-k$ 
  we obtain $  \xi_{(i)}\ -\ \xi_{(i-k)}  \ < \  k\cdot\Delta_{thr}   \  (\ k\ \leq\ N_{tol} \ )$.\\ 
  With this cluster definition, both elements $i$ and $i+k$ (or $i-k$) are elements
  of the same cluster. Moreover all the elements between $i$ and $i+k$ (or
  $i-k$) are elements of the same cluster.\\
  If $N_{tol}\ =\ 1$ the cluster definition  of eq. \ref{eqn:cluster2}
  corresponds to the definition in eq. \ref{eqn:cluster1}.
  The cluster definition of eq. \ref{eqn:cluster2}  simply states that on average the distance among
  elements of a cluster must be $\le\ \Delta_{thr}$. N$_{tol}$ is the maximum allowed
  number of elements for which the average distance can be evaluated. \\ 
  This generalization of the definition of clusters largely reduces fragmentation of periods of flat flux.

  The cluster definition of  eq. \ref{eqn:cluster2} searches for clusters
  starting from contiguous events, and two contiguous flares ($F_a$ and $F_b$) are glued together
  when the first event of $F_B$ and the last event of $F_A$ obey eq.
  \ref{eqn:cluster2}. It is necessary for gluing that there are no more than $2\cdot N_{tol}$ events in-between
  the two flares.
  The proposed procedure does not try to merge distant flares, with the cost of
  introducing the $N_{tol}$ parameter.
  Instead, the method proposed in \cite{cucala2008} try to merge distant
  flares. In appendix \ref{appendix:gluing} I will discuss the gluing effect.\\
 
\section{The iterated clustering procedure (iSRS)}

For a given ordered data-set, events can be clustered choosing $\Delta_{thr}$, and
N$_{tol}$.
For an extremely high $\Delta_{thr}$ (larger than the maximum spacing
between two contiguous events $\Delta^{max}$) the ground cluster $C_0$ is identified. It
contains all the events of the data-set.\\
A fine Scan is performed varying the value of $\Delta_{thr}$ (but for a fixed
value of N$_{tol}$): starting with $\Delta_{thr}=\Delta^{max}$;
the scan stops when only clusters of size 2 (of two events) remain.\\
No particular attention is paid to decide the step of the scan. It has been chosen an
exponentially decreasing step, 20 steps per decade. The fine-scan on $\Delta_{thr}$ is called here {\em iterated SRS} (iSRS)
clustering
\footnote{ The C language code performing the iSRS clustering is available in electronic form
  at the CDS via anonymous ftp to \url{cdsarc.u-strasbg.fr} (130.79.128.5)
  or via \url{http://cdsweb.u-strasbg.fr/cgi-bin/qcat?J/A+A/}}.\\
Lowering the threshold (e.g., the allowed distance among elements of the
cluster), smaller clusters (with respect to $C_0$) can be obtained. The event density of each new cluster is
higher than the event density of $C_0$, because the average distance among contiguous elements of each
new cluster is lower.\\
Keeping constant N$_{tol}$, each cluster is characterized by $\Delta_{thr}$,
the cluster length, cluster size, and position within the originating
segment.\\
With the iSRS clustering, all the clusters of $m$ events can be found
($\forall\ m\ \in\ (2,N]$). We obtain a set $\{C_i\}$ of clusters ordered  by the value of $\Delta_{thr}$.\\
It could happen that the same cluster is obtained for different values of $\Delta_{thr}$.
Only one among identical clusters is maintained.\\
The set $\{C_i\}$ can be organized as a single-root tree of decreasing size (of events) and decreasing length:
Suppose a cluster $C_A$ is formed for a given $\Delta^A_{thr}$. 
Decreasing $\Delta_{thr}$, we cannot obtain a cluster including events within $C_A$ and events outside
$C_A$. In fact, to obtain this sort of cluster at a $\Delta'_{thr}$, there must exists at least an event outside
$C_A$ and an event within $C_A$ that satisfy the cluster condition of
eq. \ref{eqn:cluster2}. But, this condition is never satisfied at
$\Delta^A_{thr}$, thence it cannot be satisfied at $\Delta'_{thr}$, because
$\Delta'_{thr}\ <\ \Delta^A_{thr}$.\\
On the contrary, decreasing the threshold below $\Delta^A_{thr}$, events within
$C_A$ can form shorter clusters, because the condition of eq.
\ref{eqn:cluster2} could be only satisfied for a sub-set of the
events of $C_A$\\ 
More in general, the intersection of two
clusters $C_A$ and $C_B$ coincides
with the smallest one or it is the Empty Set.
in the case $C_A\ \cap\ C_B\ \neq\ \emptyset$, the $\Delta_{thr}$
parameter is an order parameter among the two clusters:\\
\begin{equation}
\left\{
\begin{aligned}
  \label{eqn:bigcap}
  C_A\ \cap\ C_B & = & \emptyset &  \\
  & \text{or} & & \\  
  C_A \subset C_B &  &  &\text{if}\ \Delta^A_{thr} > \Delta^B_{thr}  \\
  & \text{or} & & \\
  C_A \supset C_B  & & &\text{if}\ \Delta^A_{thr} < \Delta^B_{thr}\ .\\
\end{aligned}
\right.  
\end{equation}
where  $\Delta^A_{thr}$ and  $\Delta^B_{thr}$ are the threshold used to form clusters
A and B respectively.\\
These are the only conditions needed
to build a single-root tree structure: The whole set of clusters obtained scanning on $\Delta_{thr}$
are nodes of a tree.\\
Starting from $C_0$ and ascending the tree (e.g., going in the direction of
lowering $\Delta_{thr}$) the formed clusters are characterized
by decreasing number of events, and by the property that the
boundaries of a cluster C$_j$ is contained within the boundaries of other clusters
({\em ancestor} clusters).

We can identify an ancestor/descendant hierarchy:
Starting from a cluster $C_A$, it can be identified as an ancestor if there
exists at least an other cluster $C_B$ such that:
\begin{equation}
\label{eqn:ancestor}
  C_{A} \subset C_{B}\ .
\end{equation}
If such a condition is satisfied, $C_B$ is a descendant of $C_A$; otherwise
$C_A$ is a leaf of the tree.\\
C$_0$ is the ancestor of all the other clusters. \\
Due to eq. \ref{eqn:bigcap}, the total number of built clusters is $\leq\ N$.\\

\section{Coincidence Cluster Probability}
\label{section:coincidence} 
There is a  chance that reconstructed clusters does not represent a
flaring period, but a statistical fluctuation over the true flux
of the source. To estimate the probability of obtaining a cluster by chance,
I consider as null hypothesis, the case that the whole sample is uniformly
distributed within the extraction interval (for the Fermi-LAT data sample
 the case that the background diffuse
emission, background sources and the foreground source
give a steady contribution during the observing period within the extraction
region $R_{68}(type_i,E_i)$).\\
Suppose we have $N$ uniformly distributed events within the extraction interval.
For the case of the Fermi-LAT data sample the extraction interval is the cumulative
exposure domain (it has not to be confused with the extraction region in aperture photometry);
let us assume without loss of generality
that the extraction interval is the unitary extraction interval $(0,1]$. Suppose we
count the events within a window of fixed length ($d$) within the extraction
length. Suppose also that the window is moved within the unitary
extraction interval.
Following the notation in \cite{glaz1994},
scan-statistics evaluates the probability $P\{N_d>m\}$ to found more than $m$ events
within a moving window of length $d$ (with $d$ contained in the unitary extraction length),
where:
$N_d\ =\ sup\{N_{x,x+d};0\ \leq\ x\ <\ 1\ -\ d\}$,
$N_{x,x+d}$ is the number of events in the interval $(x,x+d]$, $0 \leq d < 1$
\citep[see][for a detailed explanation]{glaz1994}.\\
In spite of the ease of the enunciation, statisticians took over 30 years
to found a solution for $P\{N_d>m\}$ \citep{huntington1975}. The
implementation of the solution is practically unfeasible and approximate
solutions are often proposed \citep[see, e.g.,][]{huffer_lin1997,haiman2009}.\\
To approach the problem I made use of the relation reported in  \cite{glaz1994}:
\begin{equation}
 P\bigl\{N_d>m\bigr\} \ = \ P\bigl\{S_{(1)}^{(m)}\bigr\}
 \label{eqn:scanstat_mspacing_rel}
\end{equation}
where P\{$S_{(1)}^{(m)}\}$ is the distribution of the smallest of the m-spacings.
The m-spacings $S_i^m$ are defined by:
\begin{equation}
S_i^m\ =\ X_{(m+i)}\ -\ X_{(i)}
\end{equation}
where $\bigl\{X_{(i)}\bigr\}$ is the ordered set of uniformly distributed events $\bigl\{X_i\bigr\}$,
and $\bigl\{S_{(i)}^{(m)}\bigr\}$ is the ordered set of m-spacings $\bigl\{S_i^{(m)}\bigr\}$
\citep[see][and references thereafter for detailed definitions]{glaz1994}.\\
The distribution of $S_{(1)}^{(m)}$ can be easily obtained with
  simulations.
Tables were prepared with the cumulative distribution of $S_{(1)}^{(m)}$ for a
uniform distribution of {\em N} points on a segment.
Tables cover values of  {\em N} from 3 up to $10^6$.\\
Tables are filled up running $N_R=4\times$10$^6$ random samples of {\em N} events extracted with a uniform distribution.\\
The number of total random extractions is of the order of
$10^{15}$ to fill all the tables. I used the Marsaglia-Zaman RANMAR random
engine  \citep{marsaglia1990,james1990} contained in the CLHEP library which has a very long recycling-period
$\sim10^{144}$, and does not show correlated sequences of extracted variables (nearby generated points are not
mapped into other sequences of nearby points).\\

Each table corresponds to a fixed sample size $N^*$. Each row of a table contains
the distribution for a fixed m-spacing. Each element of a row reports the
length $S_{(1)}^{(m)}$ of the m-spacing which corresponds to a certain probability
$P^*\bigl\{S_{(1)}^{(m)}\bigr\}$.
Columns are prepared for probabilities
$P^*=\tfrac{1}{\sqrt{2\pi}}\int_t^\infty e^{-\tfrac{t^2}{2}}dt$
with t=2, 2.5, 3, 3.5, 4, 4.5. 
The reported lengths of m-spacings have a statistical
accuracy which corresponds to a relative accuracy on probability $\tfrac{\Delta P^*}{P^*}$ of 0.3\%, 0.6\%,  1.3\%,
3.3\%, 8.8\%, 27\% for t=2, 2.5, 3, 3.5, 4, 4.5 respectively. \\
For N $\le$ 300, the tables for all the m-spacings and for all the sample
sizes were filled. \\

   Scan-statistics cannot be applied directly to search for flaring period,
   because it makes use of a  moving-window of fixed length ($d$ in the discussion
   above). The analyst must know the duration of the flare in advance, to
   choose the value of $d$.\\
   \cite{nagarwalla1996,glaz2006,cucala2008} investigated the problem.
   In particular \cite{glaz2006,cucala2008} applied 
   scan-statistics approach iterated on a set of windows of different
   length.\\   
   Once a set $\{C_i\}$ of clusters is obtained, we want to investigate if they could
   be considered or not random fluctuations from an uniform distribution
   extracted within the extraction interval.\\
   If the problem is limited to
   study the subsample of $\{C_i\}$ which consists only of clusters of fixed size,
   the problem is univariate, and the m-spacing statistics can be applied
   directly (using eq. \ref{eqn:scanstat_mspacing_rel} and the tables with the
   distribution of $S_{(1)}^{(m)}$) to evaluate chance cluster probability
   which is called here $P_{scan}\left(C_i\right)$ to underline that it is
   valid for the scan-statistic case).
   But the case we have to face with is that the cluster size is not held
   fixed: the distribution of $\{C_i\}$ is multivariate, and
   $P_{scan}\left(C_i\right)$ does not correspond to chance cluster probability.\\
   Anyway, we can report $P_{scan}\left(C_i\right)$ in Gaussian standard deviation units
   $t_{scan}^{i}$, where the info of $C_i$ are all contained in the index  $i$.\\
   From every $C_i$ we evaluate $t_{scan}^{i}$. From the  entire set $\{C_i\}$
   we obtain the set  $\{t_{scan}^{i}\}$. 
   We study the statistical distribution of 
   \begin{equation}
     \Theta\ =\ max\left\{ t_{scan}^{i} \right\}
     \label{eqn:scan_score}
   \end{equation}
   where the maximum is evaluated over all the formed clusters. $\Theta$
   is called {\em Maximum Scan Score Statistic} \citep{glaz2006}.
   $\Theta$ has an univariate distribution that can be
   studied with simulations in the case of an uniform data-set:\\
   For every simulated sample of size N, We apply the iSRS clustering procedure, and
   we obtain $\{C_i\}$ (and thence $\{t_{scan}^{i}\}$). From $\{t_{scan}^{i}\}$,
   we found $\Theta$ according to eq. \ref{eqn:scan_score}.\\
   From  a set of simulated samples (with  the same size N),
   the distribution of $\Theta$ is obtained.\\

\begin{figure}
   \centering
   \includegraphics[width=\hsize]{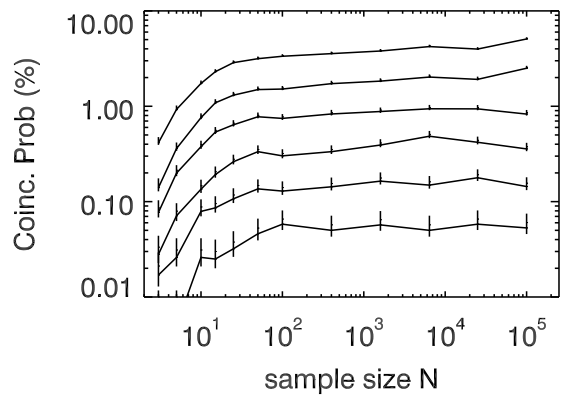}

   \caption{Cluster Coincidence Probability prepared with a Monte Carlo for samples of $N$ events.
     Curves refer  to the frequency of false-positive samples obtained for different
     $\Theta^*$ thresholds: from top to bottom $\Theta^*\ =$ 2.75,
     3, 3.25, 3.5, 3.75, 4.}

   \label{fig:corr_table}
    \end{figure}

   We define {\em false-positive} a sample with $\Theta$ above a predefined
   value $\Theta^*$. $f_{coinc}$ is the  frequency of false-positive samples.
   It can be denoted with $f_{coinc}(\Theta^*)$. The cumulative distribution
   of $\Theta$ is $\ 1-f_{coinc}(\Theta^*)$.\\
  The Monte Carlo results are reported in figure \ref{fig:corr_table}: I show the
  obtained $f_{coinc}$ as a function of sample size $N$, for a set of chosen
  $\Theta^*$.\\
  The curves reported in figure \ref{fig:corr_table} shows a
  change of slope for sample-sizes below $\sim$30.
  It is neither due to approximations on
  the distribution of $S_{1}^{(m)}$, nor to  the interpolations for the
  m-spacings and for the sample-sizes, because, for sample-sizes below
  $\sim$300, the tables for all the m-spacings and for all the sample sizes
  were filled.\\

  The m-spacings tables were used to evaluate $f_{coinc}$. 
 If a table for a  given sample-size has not been prepared, an interpolation using the tables
 with the nearest sample-sizes is performed.
 If a m-spacing row for a certain sample size has not been prepared, an
 interpolation using the nearest m-spacings is performed.
 The systematic on false-positive frequency is due to the accuracy of the
 m-spacings tables, and on the interpolations on sample-size and on m-spacing.
 Statistical accuracy of m-spacing tables has been already
 discussed. Interpolation on sample size was found to introduce a systematic
 relative error of 10\% on the evaluation of $f_{coinc}$. The interpolation on
 m-spacings introduces a systematic relative error of 7\%, 2\%, $<1$\% for
 sample of size $10^5$, 25000, 1600 respectively.\\
  In the following, I refer to the probability distribution of $\Theta$ with
  $P_{iSRS}$ ($t_{iSRS}$ in standard Gaussian units) to underline that it is
  obtained without constraints on the cluster-size, and using the iSRS
  clustering scheme.\\
  
\section{Removal of random Clusters}
\label{section:removal_random}
We have prepared a set $\{C_i\}$ from the original sample using the iSRS clustering. 
If the source was steady, the distribution of $\Theta$ prepared
applying eq. \ref{eqn:scan_score} to the set $\{C_i\}$ is known. We can
choose a confidence level, and find the corresponding $\Theta_{cl}$: If we
have an uniformly distributed sample of size N, there is a probability coincident with the confidence level that:
\begin{equation}
  \Theta\ <\ \Theta_{cl}\ .
  \label{eq:score_test}
\end{equation}
If the source, instead, had a flare, the  m-spacings are not distributed as
in the case of an uniformly distributed sample. As a consequence, $\Theta$
also is not distributed as in the case of an uniformly distributed sample, and
for a sub-set of clusters $\{\bar{C_i}\}$, we could obtain values of $t^i_{scan}\ > \Theta_{cl}$. 
I will use eq. \ref{eq:score_test} to test the hypothesis that the sample
under investigation is uniformly distributed,
and I call {\em statistically relevant} with respect to the whole investigated period the clusters of the sub-sample $\{\bar{C_i}\}$.\\

For a reason that will become obvious in a while, a special object is added as first element of
$\{C_i\}$. Its boundaries are the start and stop of the observation; Its effective length (${\tilde l}_{whole}$) is the whole extraction
interval, and its size ($N_{whole}$) is the sample size N. I will denote
it with $C_{whole}$. It does not obey eq. \ref{eqn:cluster2}. It is the {\em root}.\\

The set $\{C_i\}$ is ordered with respect to $\Delta_{thr}$.
The clusters of the ordered set $\{C_i\}$ have to be considered as candidates.
A discrimination procedure is applied in order to build a sub-set of
clusters that describe the source variability within the investigated period:\\
The removal procedure is applied to all the clusters
starting from $C_{0}$, and continuing with the clusters with lower $\Delta_{thr}$.
At first step,  we have to evaluate if $C_{0}$ has to be considered as a
random fluctuation from an uniformly distributed sample of size $N$ and
extracted within the whole period of investigation. If $C_0$ belongs to the
sub-set of statistically relevant clusters it is maintained, otherwise
it is removed from the original set.\\
If $C_0$ is maintained, the most conservative hypothesis is that the events
within $C_0$ are uniformly extracted within a period of length coincident with the
effective length of $C_0$. \\
Going on with the removal procedure, for a certain index
of the ordered list of  candidate clusters, we found
a cluster $C_{*}$.
We have to choose if accept or reject $C_{*}$:
We identify uniquely its direct accepted ancestor $C_p$ (parent). 
Due to the fact that the removal procedure is ordered with respect to
$\Delta_{thr}$, $C_p$ has already been accepted.
The most conservative hypothesis (null hypothesis) is that the sample (of size $N_p$) of the events
contained in $C_p$ is uniformly distributed within the effective length
(${\tilde l_p}$)
of $C_p$ (e.g., that the source flux was steady during the period identified
with C$_p$).
If no parent cluster exists, the root ($C_{whole}$) is considered as parent instead,
and $N_{whole}$ and ${\tilde l_{whole}}$  are used.
We accept $C_{*}$ if the null hypothesis is rejected according to the chosen
confidence level.
The removal decision of cluster $C_*$ follows the same arguments used for the
removal decision of cluster $C_0$, but instead of evaluating  if $C_{*}$ is a
statistically relevant cluster with respect to the whole investigated period,
we make the following evaluation:
We restrict to the reduced sample of events contained in $C_p$. This sample is of size
$N_p$, and it is assumed to be of length ${\tilde l_p}$.
We evaluate if $C_{*}$ is among the statistically relevant clusters for the reduced
sample. I call these clusters {\em statistically relevant with respect to the period described by the
cluster $C_p$}, or, more concisely: {\em statistically relevant with respect to $C_p$}.
If $C_{*}$ is statistically relevant with respect to $C_p$, it is maintained,
otherwise it is removed.
If $C_*$ is maintained, the new conservative hypothesis is that $C_*$ describes a period of steady
activity.\\

\begin{figure}
   \centering
   \includegraphics[width=8cm]{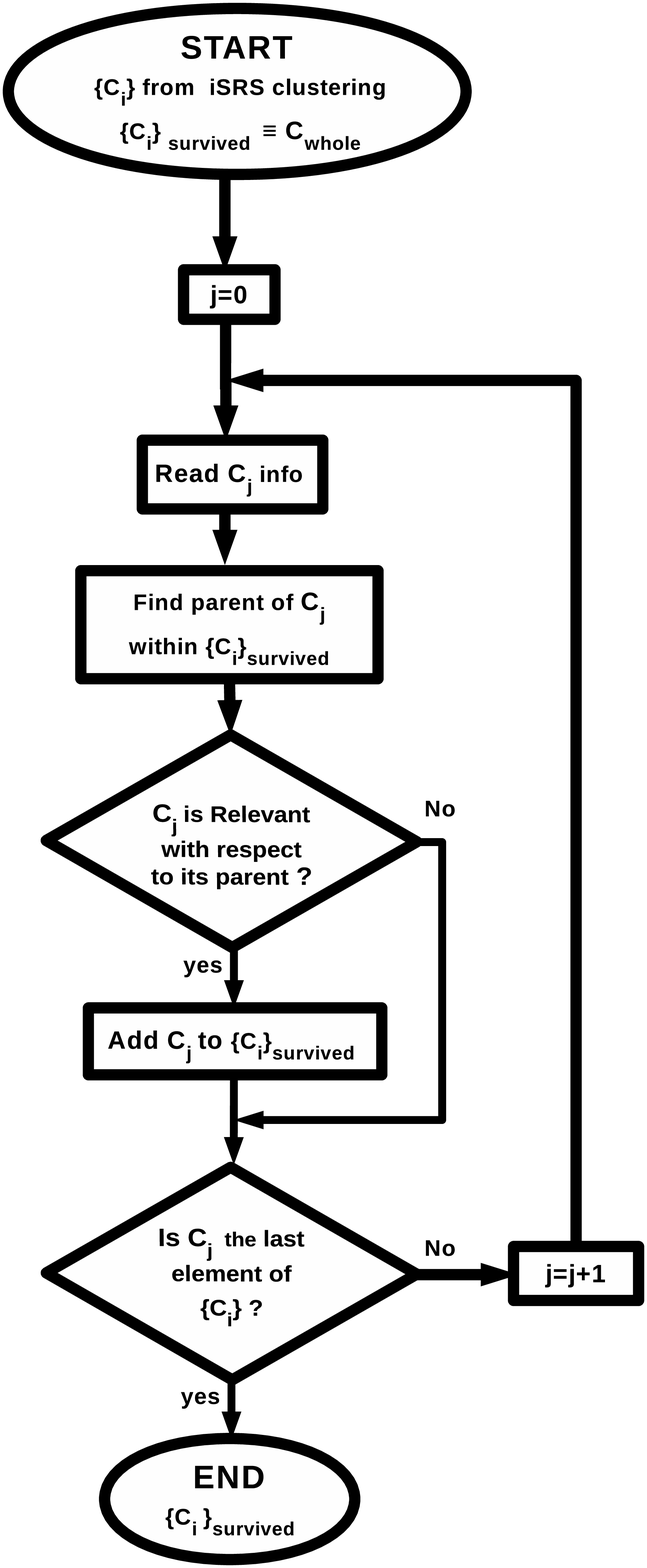}
   \caption{ 
     Flowchart of the removal of random clusters. At beginning we have
       the ordered set $\{C_i\}$ obtained with the iSRS clustering. The
       initial list of survived clusters $\{C_i\}_{survived}$ contains only
       the special cluster $C_{whole}$.       
       At step j, 
       cluster $C_j$ of the ordered set $\{C_i\}$ is evaluated: its parent is
       identified within the current list  $\{C_i\}_{survived}$ (if no parent cluster exists, $C_{whole}$ is used as parent).
       If the cluster $C_j$ is among the statistically relevant clusters with
       respect to its parent, it is added to  $\{C_i\}_{survived}$.
       When all the cluster of the initial set $\{C_i\}$ are evaluated, the
       procedure stops and $\{C_i\}_{survived}$ is built. 
   }
   \label{fig:removal_random}
    \end{figure}

The statistical discrimination procedure is performed for all of the candidate
clusters. The flowchart of the procedure is reported in
  fig. \ref{fig:removal_random}. \\
The surviving clusters have new properties:
\begin{enumerate} 
\item
They are all statistically relevant with respect to their own parents
\label{itm:Crelevant};
\item
The events contained in any parent cluster do not belong to an uniform
distribution \label{itm:Cnot_uniform}.
\end{enumerate}

\section{Unbinned light curves}
\label{section:ulc}

Two opposite scenarios are discussed: a steady, and a flaring source. \\
If the source was steady during the whole observing period, we expect no cluster to
survive the removal of random clusters, apart the root $C_{whole}$.\\
There is a chance probability  $<$1.3\textperthousand (if the confidence
level is set to $99.87\%$) to obtain one or more clusters from a uniformly
distributed sample.\\

It is useful to walk again through the iterated SRS clustering, and
  through the removal of random clusters for the case of a source that
  underwent a flare during the investigated period.
The cluster property \ref{itm:Cnot_uniform} states that clusters, and 
chains of clusters are expected during activity periods (when the hypothesis
of steady source is false):\\
Once the set $\{C_i\}$ is prepared,
there could be at least a candidate son cluster C$_i^{whole}(\xi_{thr_i})$
which is statistically relevant with respect to the whole observing
  period (if there are several relevant clusters, the one with the largest
  $\Delta_{thr}$ is chosen).
In this case, the null hypothesis of a steady source is rejected,
and the new conservative hypothesis is that C$_i^{whole}$ corresponds to a
period of flat activity.\\
Continuing the removal of random cluster procedure, 
we could find at least a cluster C$_j^i$ (descendant of an
accepted cluster $C_i$) which is statistically relevant with respect to the
period identified with $C_i$ (if there are several relevant clusters, the one with the largest $\Delta_{thr}$
is chosen). 
We reject the hypothesis of a steady source
within the period identified with $C_i$, and maintain the cluster $C^i_j$. The new
conservative hypothesis is that the cluster $C^i_j$ identifies a period of flat activity.\\
The procedure stops when no son clusters exist for which the new conservative hypothesis
can be statistically rejected.\\
Clusters with no sons identify the flare peaks:
the leaves of the three are found when the flare peak is found
(the light-curve at its peak is locally flat), or the paucity of
collected photons prevents the procedure to go on). The set C$_i^{whole}$,C$_j^i$ ,C$_k^j$, ..., C$_m^l$, C$_n^m$ is the
chain describing the flaring period.  C$_n^m$ is a leaf of the tree. It
represents the activity peak.\\
Property \ref{itm:Crelevant} states that each cluster of the chain is statistically relevant with respect to its parent: the
chain of clusters is a statistically filtered representation of the flaring period.
Examples of real unbinned light curves
are reported in fig. \ref{fig:ulc_crab} and \ref{fig:ulc_3c454p3}. Each
horizontal segment represents a cluster of the tree: it subtends the temporal
interval characterizing the cluster; its length is the effective length
of the cluster in the temporal domain; its height is the mean photometric flux
of the source within the subtended temporal interval.\\
Each reported cluster cannot be considered a random fluctuation
  (according to the chosen C.L.) from a flat activity
period identified by its parent cluster.\\
The unbinned light curve as a whole is a representation of the tree like
hierarchy. The peaks of flare activity are the clusters with no associated
sons. This means that within each identified period of activity peak, we did
not found any statistically relevant sub-set of events describing a period
of larger flux. The identification of the activity
peaks is a direct result of the unbinned light curve procedure.\\
The reported clusters are not
independent. Eq. \ref{eqn:bigcap} and \ref{eqn:ancestor}
state that clusters describing the same flare are all correlated, because
their intersection is not the empty set. Thence, the evaluation of the temporal FWHM
of reconstructed flares can be performed starting from the unbinned light curve, but
the statistical uncertainty of the temporal FWHM must be evaluated using simulations.
\\
\section{Performance of the method for the FERMI-LAT: flare detection efficiency, flare reconstruction capability}
I tested the proposed procedure with simulations.
As far as gamma-ray background varies with celestial coordinates, and chance
detection probability depends on source mean flux and background, the performance of the
method depends on the investigated source.
I focus here to the case of the Flat Spectrum Radio Quasar  3C 454.3 .\\
The  extraction region is centered on the coordinates of the source, and its radius
corresponds to the containment of 68\% of photons from the source.
Background level corresponds to the observed background for the source
(In appendix \ref{appendix:bkg_level} I will show the method to evaluate background).
I computed the FERMI-LAT exposure for  3C 454.3 with a binsize of 86.4 s from the
beginning of FERMI-LAT scientific operations, till 2015-11-16 (see appendix \ref{appendix:data_preparation}
for the details of exposure preparation).\\
I simulated ideal flares photon by photon with a temporal shape
\begin{equation}
  \label{eqn:crrc}
  F(t)\ =\ A\biggl(1-e^{-\frac{t-t_0}{\tau_1}}\biggr) e^{-\frac{t-t_0}{\tau_2}} \ .
\end{equation}
In order to reduce the total number of simulation runs, I chose $\tau_1=\tau_2=\tau$. 
Coefficients $A$ and $\tau$ are chosen to simulate flares with a given peak flux and temporal {\em FWHM}.\\ 
Ideal flares are simulated assuming a flat exposure. 
I used the computed exposure to accept or reject simulated photons from the
source and from background. The accepted photons are the final photon list.\\
The chosen threshold probability $P_{iSRS}^{thr}$ is
99.87\% (e.g., $t_{iSRS}^{thr}$ = 3).\\

\begin{figure*}
  \centering
  \includegraphics[width=\hsize]{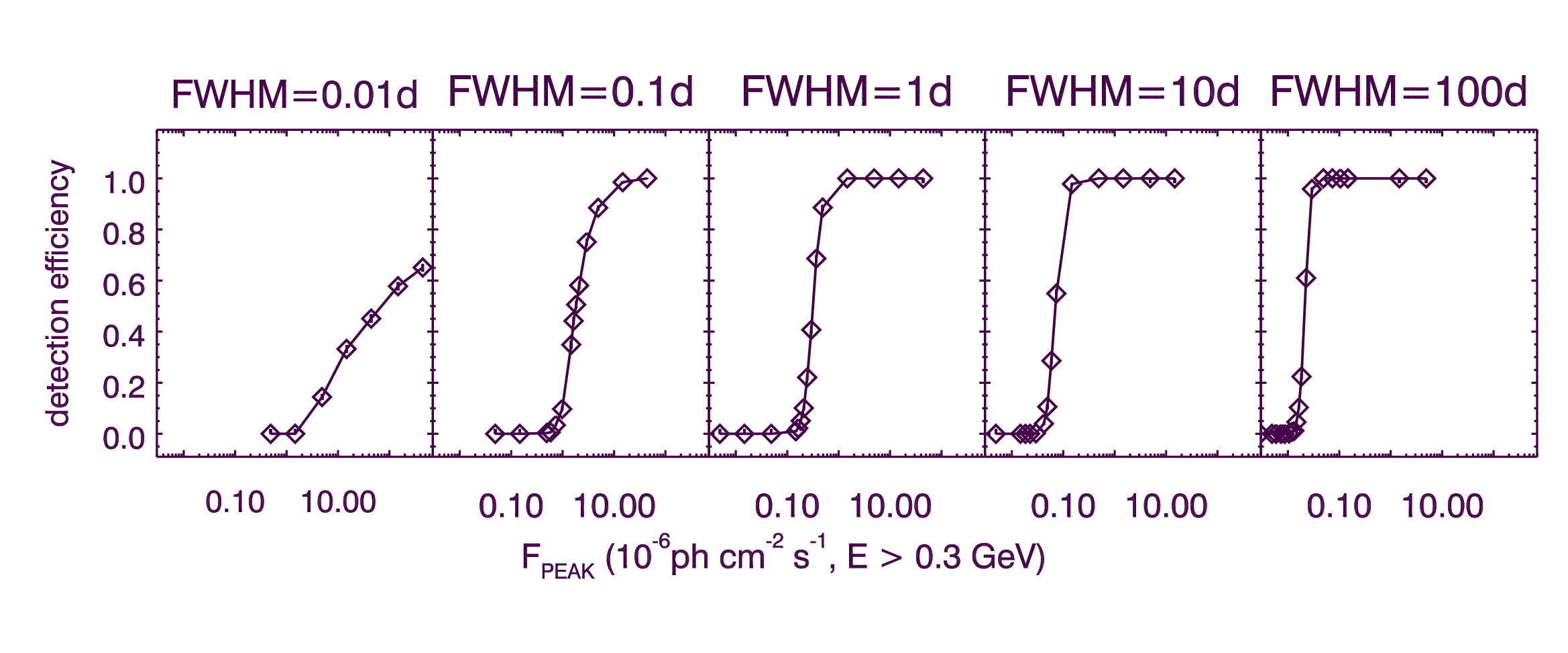}
  
  \caption{Detection efficiency for flares with FWHM of 0.01, 0.1, 1, 10, 100
    d as a function of the simulated flare peak flux. The observing period is
    7.25 y.}
  
  \label{fig:sigmoide_eff}
\end{figure*}

\begin{figure}
  \centering
   \includegraphics[width=\hsize]{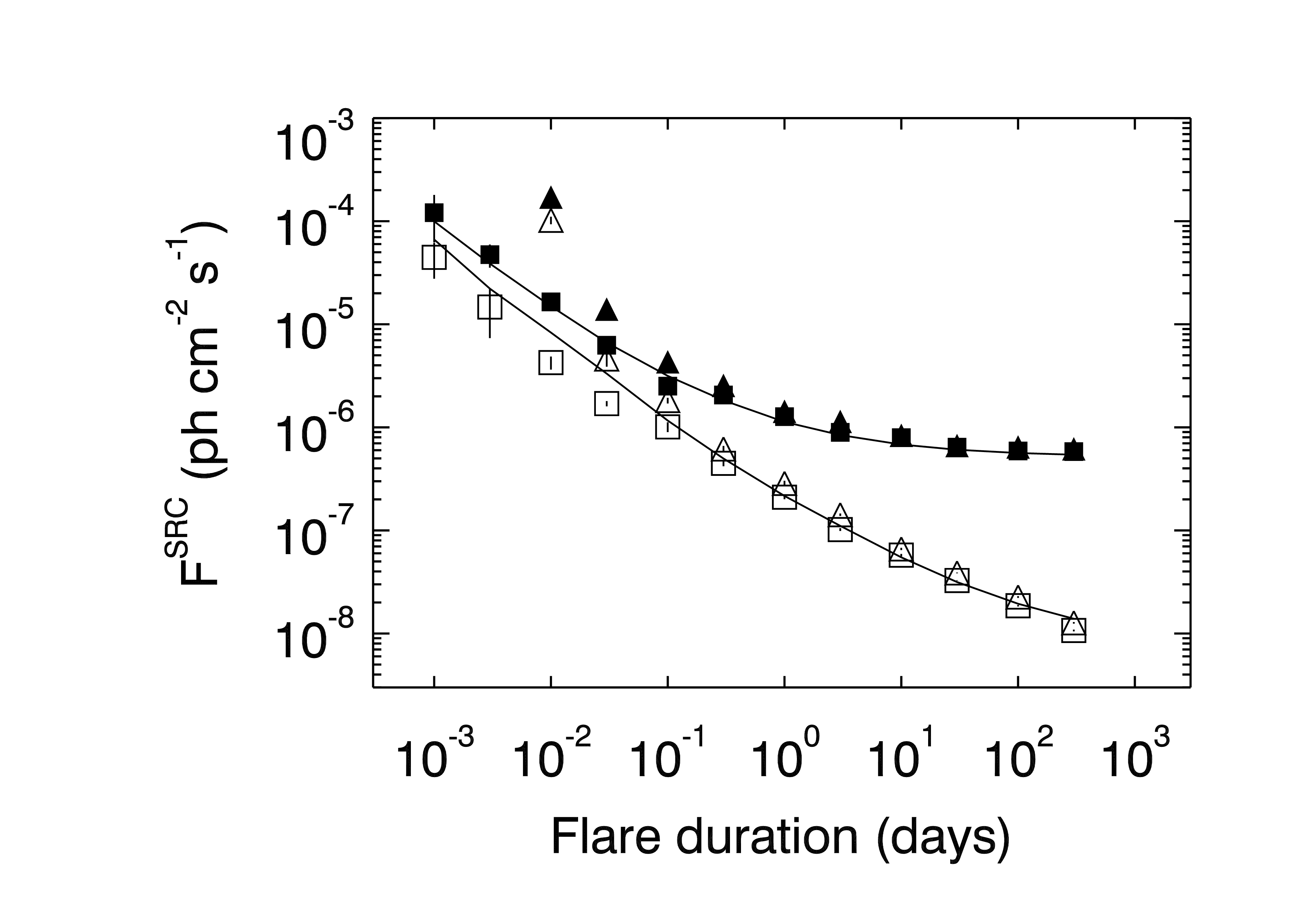}

  \caption{Peak flux detection threshold as a function of the simulated flare
    duration, reported as FWHM. Triangles (squares): Flux threshold is defined as the peak flux for
  which half (20\%) of the simulated flares are detected. 
  The observing period is 7.25 y, E $>$ 0.3 GeV, The extraction radius
  corresponds to the containment radius for the 68\% of events.  
  Open squares and triangles refers to a faint source (giving a total number of counts on
  FERMI-LAT which is 20\% of the background counts within the scrutinized period). 
  Filled symbols refer to a bright source (corresponding to the case of
  3C 454.3). 
  Curves refer to sensitivities calculated  assuming constant
  exposure, hat shaped flares, for $t_{iSRS}^{thr}\ =\ 3.$ (e.g.,
   99.87\% c.l.).}
  
  \label{fig:global_eff}
\end{figure}

In figure \ref{fig:sigmoide_eff} I report the detection efficiency for flares with a temporal
FWHM of 0.01, 0.1, 1, 10, 100 d, and with peak flux from
$10^{-8}$ to $10^{-5}$ ph cm$^{-2}$ s$^{-1}$ (E > 0.3 GeV). 
For flares with a FWHM above 1 d (slow flares), the detection efficiency rises fast around the
threshold flux. On the contrary, below 1d (fast flares), it rises
slowly, because the FERMI-LAT observes sources for windows of 10-20 minutes
each orbit or two (and sometime 4).
Extremely fast and bright flares can be detected, even if their peak
emission lies outside the observing windows, provided that the sampled tails of that flares
are bright enough to be detected.\\
The flux F$_{50\%}$ (F$_{20\%}$) corresponding to a detection efficiency of
50\% (20\%) is reported in figure \ref{fig:global_eff} as a function of temporal FWHM
of the simulated flares.
Two cases are reported, corresponding to faint ($N^{SRC}\ \le\  0.2N^{BKG}$) 
and bright sources ($N^{SRC}\ \sim 16N^{BKG}$), where
$N^{SRC}$ and $N^{BKG}$ correspond to the total source and background counts
integrated in 7.25 years with Fermi-LAT collected in the chosen extraction region.\\
Below 0.01 d, the computed values of F$_{20\%}$ is very similar for bright and faint
sources, because flares could be detected, provided
they happen outside the exposure gaps: the satellite pointing strategy affects the detection efficiency
and dominates over statistic.
\begin{figure*}
  \centering
  \includegraphics[width=\hsize]{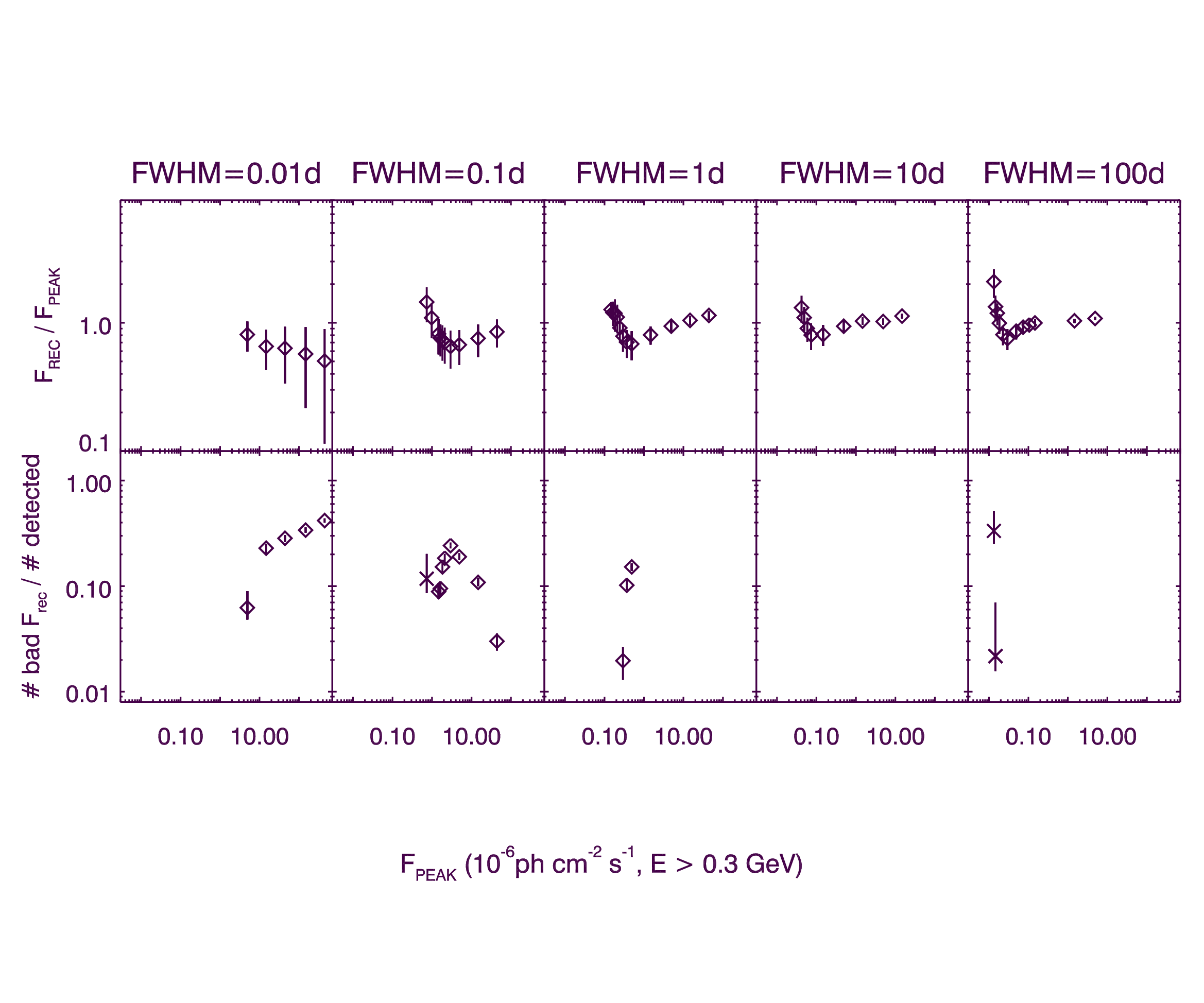}
  
  \caption{Peak flux reconstruction capability. Top panels: Mean (open diamonds) and standard
    deviation (vertical lines) of the reconstructed flux (normalized to the simulated peak flux)
    is reported as a function of the peak flux for different values of the FWHM. 
    Bottom panels: Number of flares (normalized to the number of detected flares) for which the
    reconstructed peak flux is below a factor $\tfrac{1}{2}$ of the simulated value (diamonds),
    or above a factor 2 of the simulated value (crosses).  Where values are not
    reported in bottom panels, upper limits of $6.6\perthousand$ (with 99.87\%
    c.l.) must be considered.}
  
  \label{fig:fpeak_rec}
\end{figure*}

The peak flux reconstruction capability is reported in figure \ref{fig:fpeak_rec}.
For bright flares the Reconstructed flux (F$_{rec}$) approaches F$_{peak}$.
In the region where the detection efficiency (see fig. \ref{fig:sigmoide_eff}) is larger than $\sim \tfrac{1}{2}$,
the Reconstructed flux (F$_{rec}$) is in the range F$_{peak}/2\ <$ F$_{rec}$ $\le$ F$_{peak}$.
The ratio F$_{rec}$/F$_{peak}$ increases while F$_{peak}$ increases. For the brightest flares F$_{rec}$ approaches F$_{peak}$.
The lowering of F$_{rec}$/F$_{peak}$ for faint flares is due to the fact that for faint flares, the activity peak is not well resolved.\\
In the region where the detection efficiency is smaller than $\tfrac{1}{2}$, the ratio 
F$_{rec}$/F$_{peak}$ increases while the F$_{peak}$ decreases, and it is lower than 2.\\
For flares shorter than 1 d, the reconstructed flux shows large deviations from
the simulated one, because of the FERMI-LAT observing strategy (for the
majority of the flares the peak flux is not sampled at all).
For flares of FWHM=0.01 d, on average the ratio F$_{rec}$/F$_{peak}$ 
decreases while F$_{peak}$ increases. The reason is that  faint flares are
preferentially detected when the peak of emission is sampled.\\

\begin{figure*}
  \centering
  \includegraphics[width=\hsize]{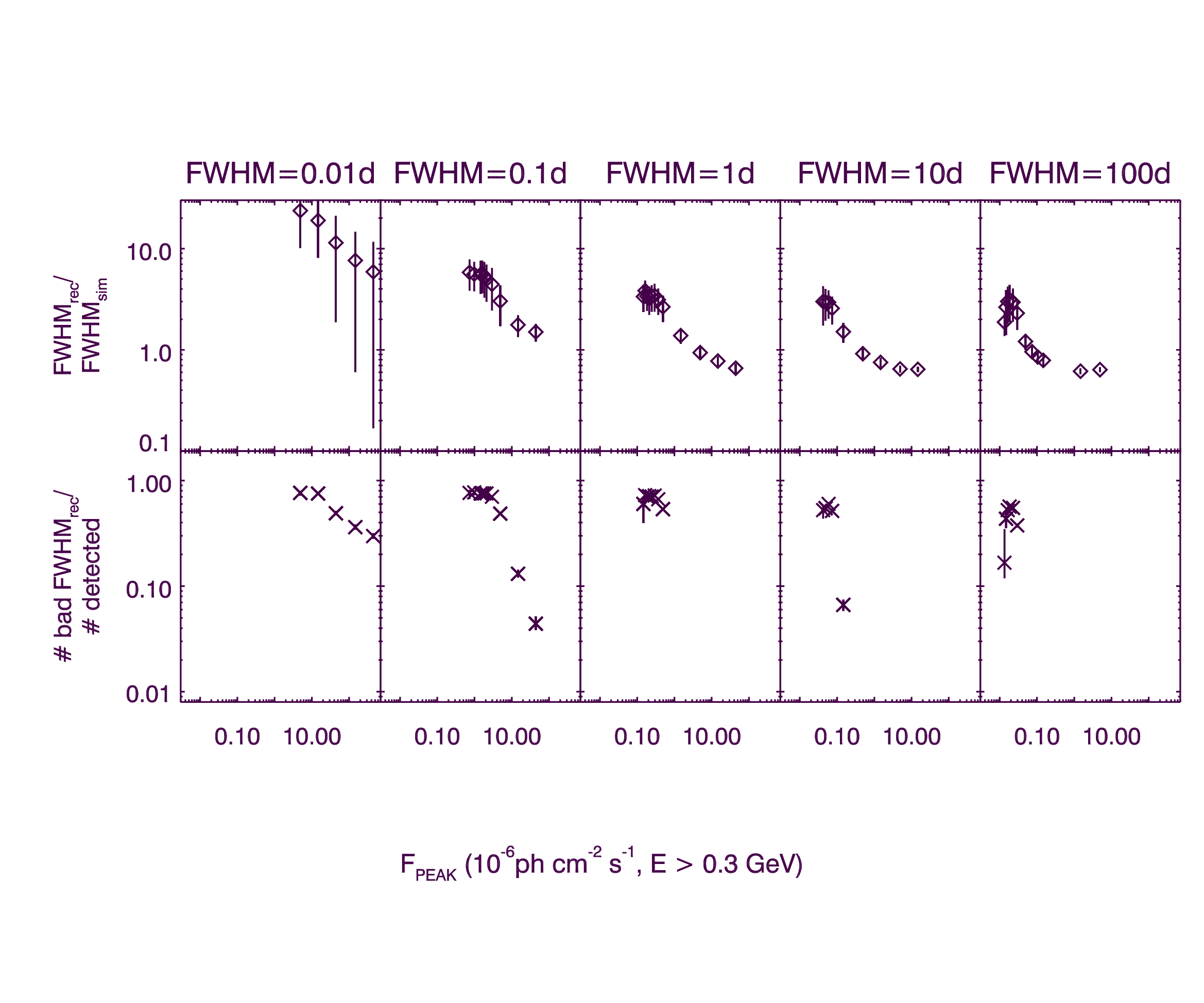}
  
  \caption{Temporal FWHM reconstruction capability. Top panels: Mean (open diamonds) and standard
    deviation (vertical lines) of the reconstructed FWHM (normalized to the simulated FWHM)
    as a function of the peak flux for different values of the FWHM.
    Bottom panels: Number of flares (normalized to the number of detected flares) for which the
    reconstructed temporal FWHM is below a factor $\tfrac{1}{2}$ of the simulated value (diamonds),
    or above a factor 2 of the simulated value (crosses). Where values are not
    reported in bottom panels, upper limits of $6.6\perthousand$ (with 99.87\%
    c.l.) must be considered. }
  \label{fig:fwhm_rec}
\end{figure*}

Temporal FWHM is estimated starting from the unbinned light curves. It is
derived assuming that the underlying flare has a pyramidal shape.
The temporal FWHM reconstruction capability is reported in
fig. \ref{fig:fwhm_rec}. For flares with temporal FWHM larger than 1 d, the
reconstruction capability is poor for faint flares, and the reconstructed FWHM
exceed the simulated FWHM.
The FWHM of flares with simulated FWHM shorter than 1 d cannot be
reconstructed, due to the temporal gaps of the exposure.

\section{Results for some gamma-ray samples and Conclusion}
Extensive results for an astrophysical problem will be shown in a forthcoming
paper.
To explain advantages and drawbacks of the method, I report here some case 
(Details about data-preparations are in appendix \ref{appendix:data_preparation}.\\
\cite{mcconville2011} studied the activity of the Flat Spectrum Radio Quasar \object{4C +55.17} from
2008 August to 2010 March. They found no evidence for gamma-ray variability
within the scrutinized period. The authors also found energy emission up to 145
GeV (observer frame). Their multiwavelength investigation shows the source is
compatible with a young radio source, with weak or absent variability.\\
The unbinned photometric light curve of the source for a
period of 7.25 years of monitoring in gamma-ray does not show flaring activity in
the range 0.3-500 GeV with a confidence level of  99.87\%.\\

\begin{figure*}
   \centering
   \includegraphics[width=\hsize]{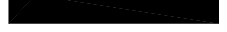}

   \caption{Unbinned light curve for the Flat Spectrum Radio Quasar  3C 454.3,
     obtained with $N_{tol}\ =\ 50$,
     $E\ > \ 0.3\ $GeV, extraction radius corresponding  to the containment of
     68\% of photons from the source,  99.87\%  c.l.}

   \label{fig:ulc_3c454p3}
    \end{figure*}

I show here the unbinned photometric light curves  for the \object{Crab Nebula} 
in the energy range 0.1-500 GeV (fig. \ref{fig:ulc_crab}), and for the Flat Spectrum Radio
Quasar  \object{3C 454.3} (fig. \ref{fig:ulc_3c454p3}) in the energy range 0.3-500 GeV.

\begin{figure*}
\centering
\begin{tabular}{cc}
\psfig{figure=Crab_uLC_nsigma3.0_55439__55481.eps,width=8cm} & \psfig{figure=Crab_uLC_nsigma3.0_55658__55675p5.eps,width=8cm} \\
\end{tabular}
 \caption{Unbinned light curve for the Crab Nebula,
           obtained with $N_{tol}\ =\ 50$,
           $E\ > \ 0.1\ $GeV, extraction radius corresponding  to the containment of
           68\% of photons from the source. Confidence level is  99.87\%.
           The unbinned light curve is produced for 7.25 y,
           but only two already studied periods are shown for a direct
           comparison of the results:
           Left panel reports the observing period investigated in
           \cite{striani2013} and in \cite{abdo2011}. Right panel  reports the
           observing period investigated in \cite{buehler2012}.
           Each horizontal segment represents a cluster: it subtends the temporal interval characterizing the
           cluster; its length is the length of the cluster in the temporal
           domain; its height is the mean photometric flux of the source
           within the subtended temporal interval.
           The unbinned light curve as a whole is a representation of
           a single-root tree like hierarchy. The bottom segment is the root
           cluster.
           Ascending the tree corresponds to go from the bottom up of the plotted diagram of clusters.
           For each cluster, a parent can be identified
           (the boundaries of a son cluster are
           within the boundaries of the parent).  
           All the reported clusters which can be regarded as
           parents do not describe flat activity periods           
           (the hypothesis that the events within a parent cluster are uniformly
           distributed is rejected with a confidence level of 99.87\%).
           Clusters and chain of clusters are expected  for flaring periods,
           when the hypothesis of uniformly distributed events is false.
           Leaves are the activity peaks.           
           Every son cluster is statistically relevant with respect to its parent,
           according to the chosen confidence level.           
           Therefore the unbinned light curve is a statistically filtered
           representation of the source activity.
 }
 \label{fig:ulc_crab}
 \end{figure*}

   These sources are among the brightest gamma-ray sources in the sky, and the
   unbinned light curves are fully representative of source variability.\\
   For both the sources I used $N_{tol}\ =\ 50$ to reduce fragmentation.
   In figures \ref{fig:ulc_crab} and \ref{fig:ulc_3c454p3} background is not subtracted, but it is negligible.
   Each horizontal line represents the mean source flux (including background
   sources and diffuse background emission).
   The ground level is the mean source flux (+background) over the whole scrutinized period.
   Each horizontal segment is statistically relevant above the father one
   according to the chosen confidence level.\\

The  Crab large variability in gamma-ray was first observed with AGILE
\citep{pittori2009,tavani2011}. They argue the variability is due to a
component with a cutoff around 0.5 GeV, so I report here the unbinned light
curve obtained for confidence levels of  99.87\%,
in the energy range 0.1 - 500 GeV. In the low energy band the
instrumental PSF of the FERMI-LAT is large, but the  Crab flux is extremely
bright and contamination from nearby sources can be neglected.  
The unbinned photometric light curves reported here refer to the
FERMI-LAT observing period only.\\
Due to the complex unbinned light curve of the Crab Nebula, 
I report in figure \ref{fig:ulc_crab} two periods only, referring to: the flaring period investigated  in
\cite{striani2013} and in \cite{abdo2011} (panel a); and the one
investigated in \cite{buehler2012} (panel b) .\\
The comparison among the unbinned light curve proposed here and previous works
is useful to show the power and weakness of the proposed method.
The top panel in figure \ref{fig:ulc_crab} shows a  flare that is
not well resolved in \cite{abdo2011}. The same flare is reported also as F7 in
\cite{striani2013}.
I obtain a peak photometric flux on MJD 55459.793. The temporal FWHM estimated
from the unbinned light curve reported here is $0.23\pm 0.12$ d (the error is
evaluated from the Monte Carlo simulations reported in
fig. \ref{fig:fwhm_rec}. The temporal FWHM could be
overestimated by a factor $\sim$2. The unbinned light curve analysis
shows that it is the shortest flare ever reported for the source so far.\\
The feature F6 in  \cite{striani2013} is not detected with the unbinned light
curve, but in the approach proposed here, clusters with low statistical
significance are disregarded.\\

The other panel in figure \ref{fig:ulc_crab} can be compared to finely
segmented light curve produced with likelihood analysis in \cite{buehler2012}.
That authors prepared a fixed exposure light curves (the binning is variable
in time, the mean bin duration is 9$^\prime$).
The analysis in  \cite{buehler2012}  makes use of
the Bayesian Block procedure for binned data to statistically evaluate variability, and
performs an exponential fitting of the rising part of the two resolved flares.\\
The method proposed in this paper reveals the same two flares, centered on MJD
55665.110 and MJD 55667.319, with temporal FWHM of $1.2\pm 0.4$ d and $1.1\pm
0.4$ d respectively, in agreement with the results reported in \cite{buehler2012}.\\

Flares from the FSRQ 3C 454.3 are intensively studied. 
The binned light curve for the first 3 years of FERMI-LAT
observations of  3C 454.3 is reported in \cite{abdo2011b} with a time bin of 1
day. The light curve for the first 5.1 years of FERMI-LAT
observations is reported in \cite{pacciani2014} with a time-bin of
4 days (E > 0.3 GeV). The light curve for the following 1 year of the source is reported in
\cite{britto2016} with 1 day time bin. A shorter period of activity is
reported with 3hr time bin in \cite{coogan2016} and in \cite{britto2016}.
The procedure proposed here detects the flares of the source with some exception:
It does not detect the secondary flare around MJD 55330 shown in
the 0.1-500 GeV light curve reported in \cite{abdo2011b}.
It is detected integrating data in the 0.1-500 GeV with a confidence level of
 99.87\%.\\

In  \cite{abdo2011b}, authors report a flare fine-structure during the brightest
activity period of the source (around MJD 55517 - 55520), and they
investigated the fine-structure fitting a model of 4 flares to the data in the
0.1-500 GeV energy range.
The method proposed here reveals a single flare in the energy range 0.3 - 500
GeV (with  99.87\%  c.l.), without  making assumptions on flare shape.
Substructures emerge integrating data in the 0.1-500 GeV (97.7\%\ c.l.).
In conclusion, for the brightest flares, the study of fine-structure making use of a fitting
strategy, is the preferred method, that has to be statistically compared with the null hypothesis of a single flare.\\

The unbinned photometric light curves  shown here for the  Crab Nebula
(fig. \ref{fig:ulc_crab}) and for the FSRQ  3C 454.3
(fig. \ref{fig:ulc_3c454p3})
show both the advantage of the procedure, both the
remaining drawbacks:\\
Peak flux activity period is resolved only when it is larger than the mean flux
(including diffuse background and nearby sources).
Weak flares from faint sources, surrounded by bright sources, cannot be resolved. This is the
case of 3C 345, whose  flaring activity \cite[see, e.g., ][]{atel2226} is washed out by the presence
of the nearby bright sources 4C 38.41, Mkn 501 and NRAO 512 at 2.2\textdegree,
2.1\textdegree, and 0.48\textdegree from the foreground source respectively.\\
Isolated flares are resolved according to sensitivity limit.\\
As far as there is no time-binning, there are no time-bin related issue.
The detection of fast flares from the Crab Nebula, and large temporal structures as in the case of 3C 454.3
is obtained without ad-hoc assumptions, or peculiar choices in the analysis.\\
There is however a resolving-power drawback. In fact there is the need to
define $N_{tol}$ to avoid fragmentation. The consequence is that consecutive flares,
whose peak activity are separated by less than $2N_{tol}$ gamma-rays could
 be merged together, as discussed in appendix \ref{appendix:gluing}; merging involves faint flares. \\

A fraction of fast and bright flares can be detected
depending on the pointing strategy of the FERMI satellite. I evaluated that
FERMI-LAT can detect flares with a temporal FWHM as short as
10-20 minutes and peak flux $\sim$10$^{-5}$ph cm$^{-2}$ s$^{-1}$ (E $>$ 0.3
GeV) with a detection efficiency of 20\% (at   99.87\%  c.l.). But this evaluation depends on the temporal
profile of the flare.
It is difficult, indeed, to evaluate peak flux and temporal FWHM of bright flares with typical
timescale less than 0.1-0.3 d.\\

\section{Discussion}

I have developed a procedure to identify activity peaks in time-tagged data.
I applied the proposed procedure to gamma-ray data only, but it can be applied to any time-tagged data-set,
or more in general to any ordered set of uniformly distributed events.\\
The basic task of the procedure is the identification of statistically relevant periods within a supposed uniformly distributed sample.
There are no fitting in the procedure, thence no fitting-related convergence issue.
I have shown with simulations and with real data, that the procedure is capable to detect activity peaks without any prior knowledge 
of their shape and temporal size.\\

There is a parameter in the procedure ($N_{tol}$) introduced to face with fragmentation artifact.  I have shown how to choose it,
and the drawback arising with its introduction. Small samples (of $10^3$ events or less) can be statistically investigated
to search for variability and  $N_{tol}$ could be safely set to 1 for these
samples.\\

The knowledge of performance and limitation of a method are crucial to apply it to any Astrophysical case. 
I explored the limitation of the procedure. I also investigated sensitivity limit and performance of the method for a specific case.\\

The procedure is based on scan-statistics, and on extensive Monte Carlo
simulations. Results of simulations are for a general use, and are contained in
short tables for the statistics of m-spacings and for the frequency of false-positive samples.\\
Even if it is conceivable to extend tables to extremely large samples, the
introduction of asymptotic scaling laws could be appropriate to overcome
large computational effort. A step toward this direction is performed in
\cite{prahl1999}. The author developed a Poisson test based on a
semi-analytical method. The test is based on the distance among contiguous
events (e.g. on the m-spacing problem, with $m=1$). The extension of that
semi-analytical method to $m>1$ is indeed mandatory to explore the full
spectrum of variability for extremely large samples.
Other approaches to obtain asymptotic solutions to the problem are investigated, e.g., in 
\cite{huffer1997} and in \cite{boutsikas2009}.\\ 
The computational effort to produce unbinned light curves from an ordered set
of $N$ events is extremely low (once the chance cluster probabilities are all
tabulated):
The procedure that produces clusters evaluates
no more than $N_{tol} \cdot N$ distances among events. For all the examples
reported, the iteration on  $\Delta_{thr}$ spans no more than 5 decades
with 20 steps per decade. The total number of evaluated distances is thence
no more than $100\cdot N_{tol} \cdot N$.\\
The total number of clusters is no
more than $N$, because the intersection of two clusters coincides
with the smallest of them or it is the Empty Set.\\
As far as chance cluster probabilities are all tabulated, no more than $N$
probabilities are extracted from the tables.

\appendix
\section{Fragmentation artifact and the choice of $N_{tol}$ parameter}
\label{section:fragmentation_and_ntol}
\begin{figure*}
  \centering
  \includegraphics[width=\hsize]{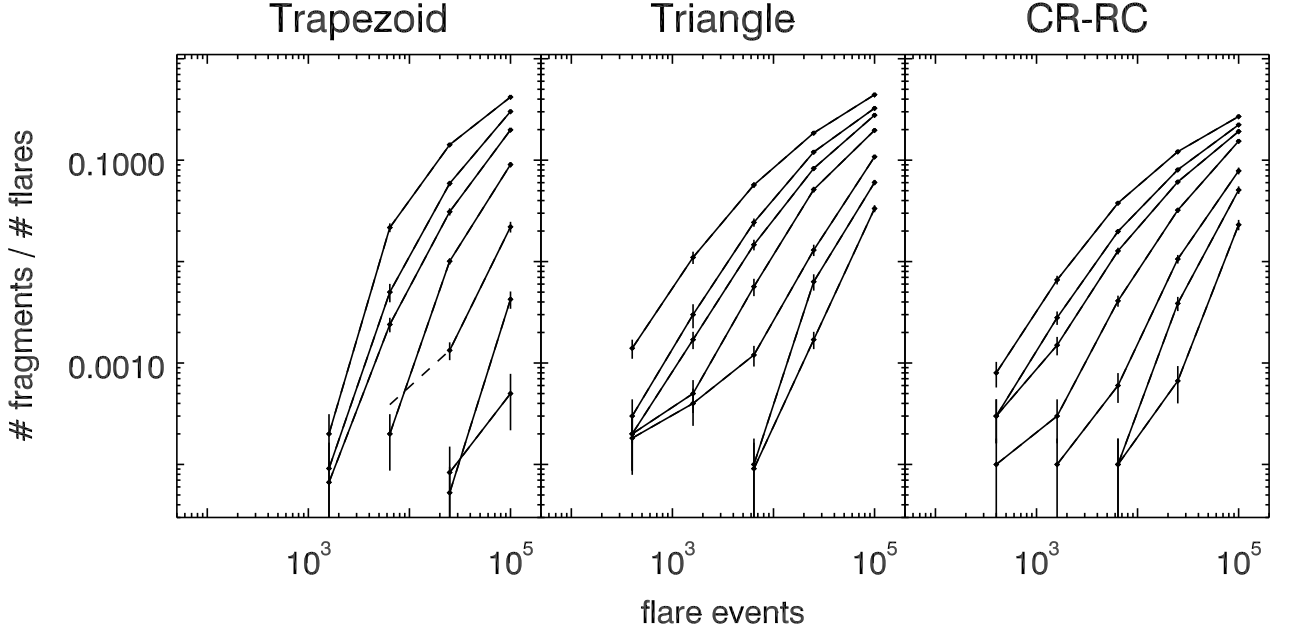}
  
  \caption{Frequency of Fragmentation artifact as a function of the events collected during the flaring period. The three panels are for the following temporal shapes:
    trapezoidal (left panel), triangular (central panel), temporal shape described in eq. \ref{eqn:crrc} (right panel).  The clustering scheme of eq. \ref{eqn:cluster2} is used. Continuous lines are used to connect the evaluated frequencies for the same tolerance parameter. Dashed lines are used to connect evaluated frequency to upper limits (99.87\% c.l.). From top to bottom curves, tolerance parameters is set to 1, 3, 5, 10, 25, 50, 100. }

  \label{fig:fragmentation_multiplot}
\end{figure*}

In the proposed procedure, fragmentation is an artifact introduced with
clustering: In the case of a period of flat source activity (with flux $F_{flat}$),
the analyst expects to find a single cluster describing the flat period.
The construction of clusters making use of eq. \ref{eqn:cluster1}
lead to the subdivision of the flat period in several clusters.
I call fragmentation this artifact. In fact, for $\Delta_{thr}\ \sim\ 1/F_{flat}$, we expect half of the
exposure spacings among contiguous events to be larger than $1/F_{flat}$. The
result is the constructions of several clusters during the flat period.\\
Fragmentation is not limited to periods of flat activity, but can occur also for flaring periods: 
If we have a flare, once we have chosen a value for $\Delta_{thr}$, we expect to build a single cluster,
but, due to the stochastic nature of the problem, several clusters can form.\\
The clustering scheme obeying eq. \ref{eqn:cluster2} reduces the occurrence of fragmentation
provided that a suitable $N_{tol}$ parameter is chosen.\\
The evaluation of the occurrence of fragmentation
is performed here with simulations, testing the proposed algorithm with three
test-functions: 
trapezoidal,  triangular, or the temporal shape described by equation \ref{eqn:crrc}.
Background, and steady source activity is neglected.
For the case of trapezoidal shape, the 80\% of events are simulated within the flat period (the superior base of the
trapezoid).  I consider the cases  that the instrument collects 400, 1600, 6400, 25$\cdot 10^3$, $10^5$ photons during the activity period
under examination.
For each temporal shape, and each photon statistic, I run up to $3\cdot 10^4$ simulations. Peaks are identified following the procedure
proposed in this paper. The tolerance parameter is set to 1, 3, 5, 10, 25, 50, 100. Fragmentation appears if
there is more than one detected peak for a simulated activity period (e.g., for a simulation).
Results are shown in fig. \ref{fig:fragmentation_multiplot}.\\
A tolerance parameter of 50 is suitable for flares with up to 25$\cdot 10^3$ events: with this choice fragmentation occurrence is at the level of 1\% or less.
Moreover this value is useful to describe large activity plateau with $10^5$
events. For flaring periods  of $10^3$ events or less, a tolerance parameter of 1
could be chosen. \\
The following procedure is suggested to correctly choose the $N_{tol}$
parameter:
Obviously, fragmentation probability rises with the number of flare events,
and it decreases while increasing $N_{tol}$.
If we could know the number of events ($N_f$)  of the activity periods, we could evaluate
the fragmentation probability at a given $N_{tol}$ and at the number of flare
events corresponding to $N_f$ for the worst case among the three test
functions reported in fig. \ref{fig:fragmentation_multiplot}.
Then we could choose $N_{tol}$ that keeps fragmentation probability at the
desired level.\\
Before data analysis is performed, we do not know the number of events of each flare, but
the total sample size is known. If the sample size is $10^3$ events or less,
the tolerance parameter can be set to
$N_{tol}\ =\ 1$. This choice allows for the fragmentation artifact to be below
1\% for each detected flare (from the worst case among the ones reported in
fig \ref{fig:fragmentation_multiplot} for $N_{tol}\ =\ 1$).\\
For larger sample sizes, the suggested procedure is to prepare a preliminary
unbinned light curve with the tolerance
parameter set to a large value (e.g., $N_{tol}\ =\ 100$). From the set $\{C_i\}$ of
clusters surviving the removal procedure, each son of the root contains all the events of one or more
flares, and they are the clusters with the largest size. The analyst must choose
the accepted cluster with the largest size (excluding the root).
The number of events of this cluster ($N^{max}$) is equal or larger than
the number of the events of the brightest flare. $N^{max}$ can be used
together with the fragmentation estimates reported in fig. \ref{fig:fragmentation_multiplot}
to choose the correct value for $N_{tol}$ in order to maintain the fragmentation
probability below a predefined level for each flaring period.\\
The obtained value of $N_{tol}$ can be used to produce the final unbinned
light curve.\\
As an example, if the analyst finds from the preliminary unbinned light curve
that the cluster with the largest size (excluding the root) contains 25$\cdot 10^3$
events, from the three test functions reported in fig.  \ref{fig:fragmentation_multiplot},
he obtains that $N_{tol}\ =\  50$ gives a fragmentation probability $\sim0.7\%$
for the worst case (triangular shape flare). Thence, if a fragmentation
probability $\sim0.7\%$ is considered acceptable by the analyst,  the final unbinned light
curve can be prepared with $N_{tol}\ =\  50$ .\\

\section{Multiple Flares resolving power} \label{appendix:gluing}
\begin{figure}
  \centering
   \includegraphics[width=\hsize]{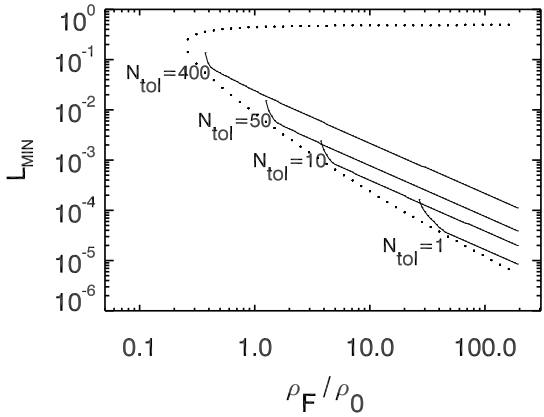}
  
  \caption{Resolving Power Capability for various $N_{tol}$
    parameters. Computation refers to two identical box shaped flares. $L$
    is the width of each flare; $\rho_F$ if the event density of the
    flare. $\rho_0$ is the background event density.
    There are $N_{tol}$ background events in between the two flares.
    The continuous curves represent the minimum flare width for which the two
    contiguous flares are resolved.
    Dotted curve is the sensitivity limit ( 99.87\%  c.l.) for
    two identical box shaped flares.
    For any given $\frac{\rho_F}{\rho_0}$ two identical flares of length lower than the
  computed values are not resolved.}
  
  \label{fig:gluing_resolving}
\end{figure}

The proposed clustering scheme (eq. \ref{eqn:cluster2}) avoids fragmentation
artifact, provided that a suitable $N_{tol}$ parameter is chosen.\\
If two contiguous flares are separated by more than $2N_{tol}$ events, the SRS
clustering scheme do not merge them. 
If two contiguous flares are separated by no
more than $2N_{tol}$ events, they could be glued together, a resolving power issue can arise:
Flares are described by a chain of clusters. 
The two flares are roughly resolved if for each flare there exists at least a
statistically relevant cluster (the cluster with the shortest $\Delta_{thr}$) for which eq.
\ref{eqn:cluster2} is not satisfied.\\
In figure \ref{fig:gluing_resolving} I show the results of such an evaluation assuming 4700
background events uniformly distributed on an unitary segment.
This background level corresponds to the background counts within a circular
region of radius $R_{68}(type_i,E_i)$ around the FSRQ 3C 454.3
 (see appendix \ref{appendix:bkg_level}).
Gluing effect is evaluated for two identical box shaped flares. The width of
each box is denoted with $L$, the event density of the flare (the flare flux) is denoted with 
$\rho_F$, the background event density is denoted with $\rho_0$.
I assumed the case in which there are  $N_{tol}$ events in between the
two box shaped flares (for a number of events larger than $2N_{tol}$   in between the two flares,
there is no gluing).
The  glued cluster can have no more than $3N_{tol}\ +\ 2(\rho_F+\rho_0) L$ events, and 
have a size $\sim \frac{3N_{tol}}{\rho_0}+2L$. I denote $C_{glued}$ this
cluster, and $\rho_{glued}$ the mean event density of the glued cluster.
I evaluated the minimum flare width ($L_{min}$) for which clusters of length $L$ and $(\rho_F+\rho_0)L$ events
are statistically relevant at  99.87\%  c.l. with respect to $C_{glued}$.
The minimum $L$ is reported in figure \ref{fig:gluing_resolving} as a function of $\frac{\rho_F}{\rho_0}$.
For comparison, I report in figure \ref{fig:gluing_resolving} also  the sensitivity limit for two identical
box-shaped flares. For a single flare, sensitivity limit has an horizontal asymptote at
$L\ =\ 1$. For two identical flares (such as in the present case), the
horizontal asymptote is at $L\ =\ 1/2$ (top sensitivity limit curve). \\
For any given $\frac{\rho_F}{\rho_0}$, two identical flares are resolved
until their width is larger than $L_{min}$. In the interval between $L_{min}$ and the
bottom sensitivity limit, the flares are detected as a single flare.
There is a range of values of $\frac{\rho_F}{\rho_0}$
for which $L_{min}$ is not defined. Within this range, the two identical flares are
always resolved, if they are detectable above background.\\
The lower $N_{tol}$ is set, the lower gluing is. The case $N_{tol}=1$ corresponds to
the simple clustering of eq. \ref{eqn:cluster1}.

\section{Fermi-LAT Data Preparation}\label{appendix:data_preparation}
I performed data preparation and likelihood analysis tasks using the standard Fermi Science Tools (v10r0p5),
the PASS8 Response Functions (P8R2\_SOURCE\_V6), and standard analysis cuts: The {\em gtselect} task was used to select
SOURCE class events (evclass=128), collected within 20$\degree$ from the investigated source.  Earth's limb gamma-rays were rejected applying a zenith angle cut of 90$\degree$. 
To prepare input files to the likelihood procedure, only good-quality data, taken during standard data taking mode, were selected using the  {\em gtmktime} task.
Livetime cube  was prepared taking into account the zenith angle cut. Exposure maps was prepared using standard recipes.\\
Unbinned likelihood analysis was performed including all the sources from the third Fermi-LAT catalog \citep{acero2015} within the chosen region of interest (ROI).\\
For the investigated source and for sources within 10$\degree$ from the ROI center, all the spectral parameters were allowed to vary. For all the other sources, only
the normalization factor were allowed to vary.\\
To prepare input files to the gtexposure and to the SRS clustering procedure,
GTIs were prepared with gtmktime task for good-quality data, taken during standard data taking mode,
and for source events outside the Earth limb.
The finely binned exposure with a binsize of 86.4 s was prepared with standard gtexposure tool, using the 3FGL catalog spectral template
for the investigated source.\\
The filtered photon list is sorted with respect to time.
The finely binned exposure, and the filtered photon-list are the inputs of the task of SRS clustering.

\section{Background Level estimate}
\label{appendix:bkg_level}
Between MJD 55800 and MJD 56100 the FSRQ  3C 454.3 underwent a period of extremely faint activity
\cite[see light curve in ][]{pacciani2014}. The unbinned likelihood analysis performed above 0.3 GeV
with a Region of Interest (ROI) of 20$\degree$ for that period gives a source flux of
$(0.75\pm 0.13)\ 10^{-8}$ ph cm$^{-2}$ s$^{-1}$, 184 source events collected by the FERMI-LAT (Npred),
and a source test statistics (ts) of 79.\\
The exposure  to the source (evaluated using gtexposure) for that period is 2.5$\cdot 10^{10}$ cm$^2$ s (E $>$ 0.3 GeV).\\
There are 704 gamma-ray events collected above 0.3 GeV within $R_{68}(type_i,E_i)$ (defined in section \ref{section:clustering_method})
for the studied faint activity period.
So I can assume $579\pm45$
 background gamma-ray within $R_{68}(type_i,E_i)$.
The FERMI-LAT exposure to the source for the 7.25 y period is 2.2$\cdot 10^{11}$ cm$^2$ s (E $>$ 0.3 GeV). The extrapolated background for the whole period
is $5200\pm400$ counts.
\\
Second method (multiple background regions method):\\
Background counts can be evaluated using a photometric method, using a suitable extraction region.
But the background level varies with galactic coordinates, so I choose to extract background counts from 36 circular background regions.
The centers of the circles are placed on a circumference of radius $R_c$=10$\degree$ centered on the position of the investigated source (3C 454.3).
They  are equally spaced on that circumference. the radius of circular background regions is dynamically chosen with the same criteria
used for the definition of $R_{68}(type_i,E_i)$, but rescaled by a factor $f_{rescale}$=4 with respect to $R_{68}(type_i,E_i)$.\\
For each background circular region, we can identify an other background region which is located in the opposite direction in the reference frame centered on the
investigated source.
I refer these two  as homologous regions.\\
The exposure of each circular background region differs to the other regions and to the exposure of the investigated source.
The sky scanning strategy of the FERMI satellite mitigates differences. Moreover, the choice of multiple background regions further reduce differences
in the effective area among the source region and the  mean of the surrounding background regions.
The usage of homologous regions brings to cancellation of differences.\\
The preparation of Fermi-LAT data for the SRS clustering handles Earth occultation, and discriminates Earth-albedo gamma-rays using 
zenith angle cuts in gtselect (see Cicerone web pages
%
\footnote{
https://fermi.gsfc.nasa.gov/ssc/data/analysis/documentation/Cicerone/\\
Cicerone\_Likelihood/Exposure.html\\
(section: {\em Excluding Atmospheric Background Events});\\
https://fermi.gsfc.nasa.gov/ssc/data/analysis/scitools/\\
aperture\_photometry.html;\\
https://fermi.gsfc.nasa.gov/ssc/data/analysis/scitools/\\
data\_preparation.html.}
).\\
The gtmktime procedure is used to prepare GTIs to account for the occultation of the source region only. The satellite is in near equatorial orbit, therefore the Earth albedo and Earth occultation GTIs  mainly depend on the celestial declination of each region. 
There are periods in which the background regions are not occulted by the Earth but  are removed from the analysis with gtmktime
(this fact gives no systematic in the evaluation of background). 
 Conversely, there are periods in which the background regions are occulted by the Earth  (or collect Earth-albedo gamma-rays) and the Earth-albedo gamma-ray
are filtered out with the gtselect procedure,
 but this filtering is not accounted for with gtmktime (gtmktime is used to prepare the GTIs to filter-out periods for which the source is below the Earth limb).
 This fact is not taken into account in the background evaluation procedure.
It gives a systematic in the evaluation of background. There are two extreme cases:
The first is for background regions located at the same celestial declination of the source,
and 10$\degree$ apart in right ascension. for such background regions, the reduction of background counts
is a factor $\frac{10\degree}{360\degree}\ \sim 2.8\%$. The detailed
semi-analytical calculation, taking into account the satellite path along the Earth, 
shows that the average loss of background events (not accounted for in the built GTIs and in the exposure calculation) is  $\sim 1.7\%$.\\
The second case is for regions at the same right ascension of the source, and the source at the edge of the Earth limb.
In such a situation, there is a loss of background counts for the regions at the lower declination that is not accounted for by the exposure. The 
detailed calculation shows that the largest loss of events among regions is $22\%$, but the loss of counts involves less than half of the background regions.
The average loss of background counts is $8.6\%$.\\
For the purpose of this paper, I disregard these systematics.\\
With this approximation and in the ideal case of no contaminating sources giving counts within the circular background regions,
the background estimated performing the 
average of two homologous regions is a linear interpolation of the background level at the position of the investigated source.\\
In order to remove contaminating sources, the three regions with the largest counts, and their homologous are neglected. 
Moreover, the three regions with the shortest counts, and their homologous are neglected. 
The background level at the position of the investigated source is estimated as the average from the counts of the survived background circular regions (rescaled by a factor $f_{rescale}^2$).\\
Using this method, the background counts within a region of radius $R_{68}(type_i,E_i)$ centered at the position of 3C 454.3,
during the extremely faint activity period, are: 540$\pm$30\ . This estimate is comparable
with the first method. The extrapolated background for the 7.25 y period is $4880\pm270$ counts.\\
Using the multiple background regions method for the 7.25 y period,
the background level within a region of radius $R_{68}(type_i,E_i)$ centered on the position of 3C 454.3, is 4680$\pm 80$.\\
The method of multiple background regions cannot be used for sources too close to the galactic plane, because in this case, the average of
the counts within the background regions is not representative of the background at the position of the investigated source.\\

\begin{acknowledgements}
I am grateful to  I. Donnarumma and A. Stamerra for
discussions. L.P. acknowledges contribution from the grant INAF SKA-CTA.
 
\end{acknowledgements}


\begin{thebibliography}{}
  
\bibitem[Abdo (2011a)]{abdo2011}Abdo, A. A., et al. 2011 Sci, 331, 739

\bibitem[Abdo (2011b)]{abdo2011b}Abdo, A. A., et al. 2011 ApJ, 733, L26
  
\bibitem[Acero (2015)]{acero2015}Acero F., et al. 2015, ApJS, 218, 23

\bibitem[Ackerman (2013)]{ackermann2013}Ackermann, M. et al, 2013, ApJ, 771, 57

\bibitem[Atwood (2009)]{atwood2009}Atwood, W. B., et al. 2009, ApJ, 697, 1071

\bibitem[Bateman (1948)]{bateman1948}Bateman, G., 1948, Biometrika, 35, 97

\bibitem[Britto (2016)]{britto2016}Britto, R. J., Bottacini, E., Lott, B., Razzaque, S., Buson, S. 2016, ApJ, 830, 162

\bibitem[Boutsikas (2009)]{boutsikas2009}Boutsikas, M. V., Koutras, M. V., Milienos, F. S. 2009, Scan Statistics - Methods and Applications, ed Glaz, J., Pozdnyakov, V. \& Wallenstein, S  Birk\"{a}user (Boston), 55
  
\bibitem[Buehler (2012)]{buehler2012}Buehler, J. D. et al, 2012, ApJ, 749, 26

\bibitem[Bulgarelli (2012)]{bulgarelli2012}Bulgarelli, A., et al, 2012, AA, 538, 63

\bibitem[Coogan (2016)]{coogan2016}Coogan, R. T., Brown, A. M., Chadwick, P. M. 2016 MNRAS, 458, 354

\bibitem[Cucala (2008)]{cucala2008}Cucala, L 2008 Biometrical Journal, 50, 299

\bibitem[Glaz (1994)]{glaz1994}Glaz, J., Naus, J., Roos, M., Wallenstein, S. 1994, Journal of Applied Probability, 31, 271

\bibitem[Glaz \& Zhang (2006)]{glaz2006}Glaz, J. \& Zhang, Z., 2006, Statistics \& Probability Letters, 76, 1316 
    
\bibitem[Haiman \& Preda (2009)]{haiman2009}Haiman, G. \& Preda, C. 2009, Scan Statistics - Methods and Applications, ed Glaz, J., Pozdnyakov, V. \& Wallenstein, S  Birk\"{a}user (Boston), 179

\bibitem[Huffer (1997)]{huffer1997} Huffer, F., 1997, Journal of the American Statistical Association (JASA), 92, 1466

\bibitem[Huffer \& Lin (1997)]{huffer_lin1997}Huffer, F., \& Lin, C. T. 1997, Journal of the American Statistical Association (JASA), 92, 1466
  
\bibitem[Huntington \& Naus (1975)]{huntington1975}Huntington R. J. \& Naus, J. 1975, The Annals of Probability, 3, 894

\bibitem[James (1990)]{james1990}James, F. 1990, Comput. Phys. Commun., 60, 329 
  
\bibitem[Lott (2012)]{lott2012}Lott, B., Escande, L., Larsson, S., \& Ballet, J. 2012, A\&A, 544, 6

\bibitem[Marsaglia (1990)]{marsaglia1990}Marsaglia, G., Zaman, A., Tsang W. W., 1990, Stat. Prob. Lett. 9, 35 
  
\bibitem[Mattox(1996)]{mattox1996}Mattox, J. R. et al. 1996, ApJ, 461, 396 

\bibitem[McConville (2011)]{mcconville2011}McConville, W. F. et al. 2011 ApJ, 738, 148 

\bibitem[Nagarwalla (1996)]{nagarwalla1996}Nagarwalla, N, 1996, Statistics in Medicine, 15, 845

\bibitem[Nauss (1965)]{nauss1965} Nauss J. I., 1965, J. Amer. Statist. Assoc., 60, 532

\bibitem[Nauss (1966)]{nauss1966} Nauss J. I., 1966, Technometrics, 8, 493

\bibitem[Pacciani (2014)]{pacciani2014}Pacciani, L. et al. 2014, ApJ, 790, 45

\bibitem[Pittori (2009)]{pittori2009} Pittori, C. et al. 2009, AA, 506, 1563

\bibitem[Prahl (1999)]{prahl1999}Prahl, J. , arxiv: 9909389

\bibitem[Reyes (2009)]{atel2226}Reyes, L. C. and Cheung C. C., 2009, ATel \#2226

\bibitem[Scargle (1998)]{scargle1998}Scargle, J. D., 1998, ApJ, 504, 405

\bibitem[Scargle (2013)]{scargle2013}Scargle, J. D., Norris, J. P., Jackson B., Chiang J.,  1998, ApJ, 504, 405

\bibitem[Striani (2013)]{striani2013}Striani, E. et al., 2013, ApJ, 765, 52

\bibitem[Tavani (2009)]{tavani2009}Tavani, M., et al. 2009, A\&A, 502, 995

\bibitem[Tavani (2011)]{tavani2011}Tavani, M., et al. 2011 Sci, 331, 736
  
\bibitem[Wallenstein(2009)]{wallenstein2009}Wallenstein, S 2009, Scan Statistics - Methods and Applications, ed Glaz, J., Pozdnyakov, V. \& Wallenstein, S Birk\"{a}user (Boston), 2 
  

\end{thebibliography}
\end{document}